\documentclass{iaus}
\usepackage{graphicx}

\newcommand{\mlu}{\mbox{$M_{\odot}$\,yr$^{-1}$}}
\newcommand{\msun}{\mbox{$M_{\odot}$}}

\title[IAUS 262.~~IR observations of nearby galaxies] %% give here short title %%
{Revealing  infrared populations of nearby galaxies using the Spitzer Space Telescope}

\author[M.~Matsuura]   %% give here short author list %%
{Mikako~Matsuura$^{1,2}$
 }

\affiliation{$^1$UCL-Institute of Origins, Department of Physics and Astronomy, University College London, 
             Gower Street, London WC1E 6BT, United Kingdom
 \\ email: {\tt mikako@star.ucl.ac.uk} \\[\affilskip]
$^2$UCL-Institute of Origins, Mullard Space Science Laboratory, University College London, Holmbury St. Mary, Dorking, Surrey RH5 6NT, 
             United Kingdom 
             }

\pubyear{2009}
\volume{262}  %% insert here IAU Symposium No.
\pagerange{1}
\jname{Stellar Populations - Planning for the Next Decade}
\editors{ Gustavo R. Bruzual \& Stephane Charlot, eds.}
\begin{document}

\maketitle

\begin{abstract}

Due to their brightness  in infrared,  asymptotic giant branch (AGB) stars are in important evolutionary
stage to be understood at this wavelength. In particular, in next decades,  when the infrared optimised telescopes,
such as the JWST and the ELT are in operation,  it will be essential to include the AGB phase more 
precisely into the population synthesis models.
However, the AGB phase is still one of the remaining major problems in the stellar evolution.
This is because the AGB stellar evolution is strongly affected by the mass-loss process from the stars.
It is important to describe mass loss more accurately so as to incorporate it into stellar evolutionary models.
Recent observations using the Spitzer Space Telescope (SST) enabled us to make a significant
progress in understanding the mass loss from AGB stars.
Moreover, the SST large surveys contributed to our understanding 
of the role of AGB stars in chemical enrichment process in galaxies.
Here we  present the summary of our recent progress.

\end{abstract}

\firstsection % if your document starts with a section,

 \vspace*{-0.6 cm}

\section{Introduction}

%==============
\begin{figure}[b]
 \vspace*{-0.6 cm}
\begin{center}
\rotatebox{270}{ 
\begin{minipage} {8cm}
% \includegraphics[width=3.4in]{mag824_8.ps} 
% \vspace*{-0.8 cm}
%\resizebox{\hsize}{!}{\includegraphics*[11,20][523,679]{mag38_8_iso.ps}}
%\resizebox{\hsize}{!}{\includegraphics*[11,20][523,679]{mag38_8_iso_massloss_low.ps}}
\resizebox{\hsize}{!}{\includegraphics*[87,155][524,717]{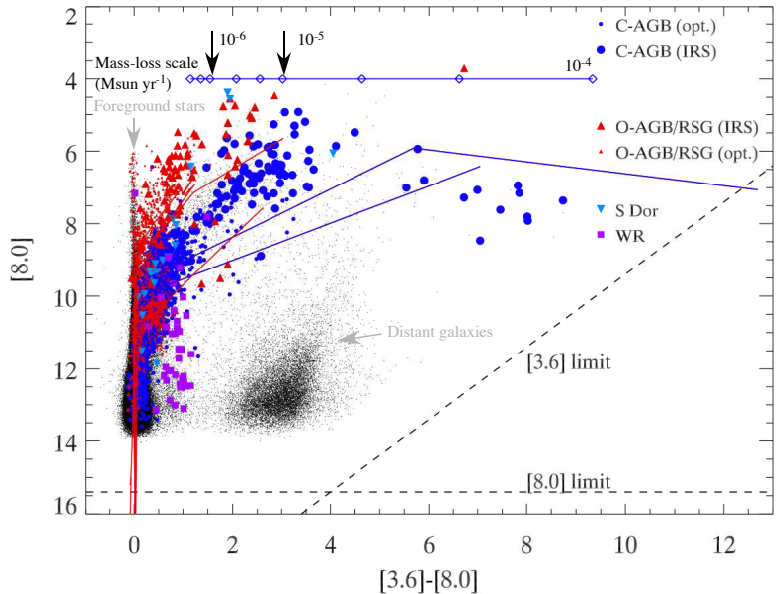}}
\end{minipage}}
 \vspace*{-0.2 cm}
 \caption{The infrared CMD of the LMC ([3.6] and [8.0] represents 3.6\,$\mu$m and 8.0\,$\mu$m magnitude 8.0\,$\mu$m, respectively).
 Dots represent the all point sources in SAGE catalogue.
 Circles show the spectroscopical identified carbon-rich AGB stars (small and large symbols show based on optical and  Spitzer/IRS spectra, respectively),
 upper triangles show oxygen-rich AGB stars and red supergiants (RSGs).
 In addition to the AGB stars, S\,Dor type variables (luminous blue variables),
 and Wolf Rayet (WR) stars are plotted.
 The dashed lines show the expected detection limit of the final SAGE product \cite[(Meixner et al. 2006)]{Meixner06}.  
 The line at the top shows the mass-loss rate vs colour relation for carbon-rich AGB stars, based on detailed analysis of 43 stars
 in the Large and the Small Magellanic Clouds \cite[(Groenewegen et al. 2007; Matsuura et al. 2009)]{Groenewegen07, Matsuura09}.
 Grey lines are AGB isochrones taken from \cite[Marigo et al. (2008)]{Marigo08}.
  }
   \label{fig1}
\end{center}
\end{figure}

%==============

%==============
\begin{figure}[b]
 \vspace*{-0.6 cm}
\begin{center}
%\rotatebox{90}{ 
%\begin{minipage} {11cm}
% \includegraphics[width=3.4in]{mag824_8.ps} 
% \vspace*{-1.0 cm}
%\resizebox{\hsize}{!}{\includegraphics*[11,20][523,679]{mag38_8_iso.ps}}
%\resizebox{\hsize}{!}{\includegraphics*[11,20][523,679]{mag38_8_iso_massloss_low.ps}}
\resizebox{0.72\hsize}{!}{\includegraphics*[15, 230][581, 605]{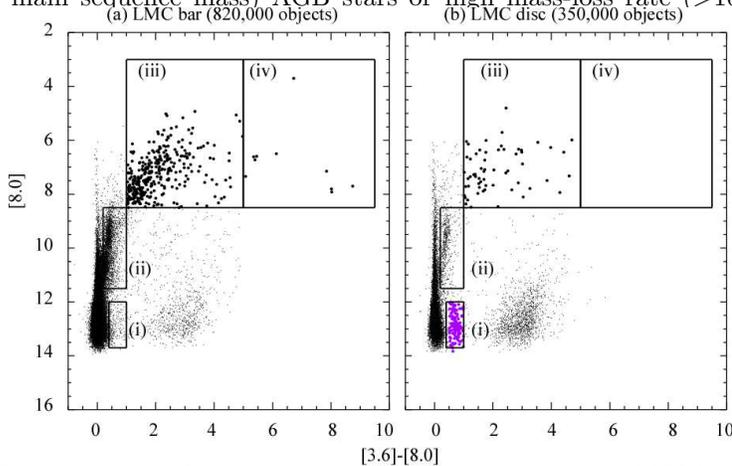}}
%\end{minipage}}
 \vspace*{-0.5 cm}
 \caption{ The comparison of the colour-magnitude diagrammes in the LMC bar (a) and north disc (b).
To emphasise the difference in these two CMDs, larger symbols are used in square regions (iii) and (iv).
 }
   \label{fig2}
\end{center}
\end{figure}

%==============

%_________________________________________________________________
\begin{table}
\vspace*{-0.6 cm}
% \centering
  \caption{ Gas and dust injected into the ISM of the LMC (from \cite[Matsuura et al. 2009]{Matsuura09})}
\begin{center}
 \begin{tabular}{lrrrrrrrrccccccccc}
  \hline
Sources & Gas mass & Dust mass & Chemical type &  \\
 & ($10^{-3}$\,\mlu)& ($10^{-6}$\,\mlu)& of dust \\ \hline
AGB stars \\
~~Carbon-rich                               & 8.6 (up to 20.6$\S$) & 43 (up to 100) & C-rich\\
~~Oxygen-rich                               & $>>$2$\dag$   & $>>$0.4$\dag$ & O-rich \\
Type II SNe                                               &  20--40 & 0.1--130$\ddag$~ & both O- and C-rich \\
WR stars                                       & 0.6 & & (C-rich?)\\
%OB stars                                        & 200--2000? \\
Red supergiants                         & $>$1                 & $>$2~~ & O-rich \\
\hline
\end{tabular}
\label{table1}
\end{center}
%$\S$ Considering the unknown dust condensation temperature, mass-loss rate can 
%be underestimated
%by a factor of up to  2.4 (Sect\,\ref{colour-mass-loss}) \\
%$\dag$ The number listed is the lower limit obtained from the observations.
%Expected gas and dust mass-loss rate are $8.6\times10^{-3}$\,\mlu and $1.2\times10^{-5}$\,\mlu, respectively. \\
%$\ddag$ Dust production (or possibly destruction) in and around SNe remains 
%uncertain.
% \gmas : gas mass-loss rates estimated using  $\Ks-[8.0]$ and $[3.6]-[8.0]$
\end{table}
%________________________________________________________________

%It will be important in next decade.
%Near-infrared regime
%Mass-loss rate is a key.
%photometry
%and
%chemical evolution
%Recent development of I
%Near-infrared spectral library

The next generation of large telescopes, such as the JWST and the ELT have many instruments optimised to near- and mid-infrared observations.
The population synthesis at these wavelength ranges will be more important.
At near-infrared $K$-band, asymptotic giant branch (AGB) stars and red giants (RGs) are 
dominant contributors to the integrated energy of $>8$ Gyr old galaxies
\cite[(Bruzual et al. 2009)]{Bruzual09}. Thus, understanding AGB stellar evolution will increase in importance.
The major uncertainty during the AGB phase is that its evolution is strong affected by mass loss from the stars,
however, the mass loss process is poorly understood. 
Here, we report our resent progress in understanding mass-loss process of AGB stars, as well as the consequence,
i.e. the importance of AGB stars on the chemical evolution of galaxies, using the Spitzer Space Telescope (SST; \cite[Werner et al. 2004]{Werner04}).

\section{Observations and results}

The SST had intensively observed very nearby galaxies,  both in photometric and spectroscopic modes.
The legacy project, SAGE \cite[(Meixner et al. 2006)]{Meixner06} surveyed the Large Magellanic Cloud (LMC) with 6 photometric bands.
We  have obtained 5--35\,$\mu$m spectra of AGB stars, using the infrared spectrometer (IRS; \cite[Houck et al. 2004]{Houck04}).
In this contribution, we focus on the results of the LMC \cite[(Zijlstra et al. 2006)]{Zijlstra06}, 
but we have also observed AGB stars in the Small Magellanic Cloud \cite[(Lagadec et al. 2007)]{Lagadec07},
the Fornax dwarf spheroidal (dSph) galaxy \cite[(Matsuura et al. 2007)]{Matsuura07}, and the Sculptor dSph galaxy \cite[(Sloan et al. 2009)]{Sloan09}.
In addition, there is a legacy program of the IRS LMC spectral survey \cite[(Kemper et al. 2009)]{Kemper09}, and these data will be available soon.
The advantage of studying these nearby galaxies is that the distances to the galaxies are used as a proxy to the distance to the AGB stars within these galaxies.
Luminosities of stars are well determined for extra-galactic AGB stars; this approach is not possible for the Galactic AGB stars.

{\underline{\it Classifications}} 
 The first step to understand the infrared survey data is to interpret the infrared colour magnitude diagrammes (CMDs).
Fig.\,\ref{fig1} shows the CMD of the LMC.
We have over-plotted AGB stars whose chemical types (oxygen-rich or carbon-rich atmosphere) are known based on spectroscopic observations.
We also plotted Wolf Rayet (WR) stars and S\,Dor variables (luminous blue variables).
The details of the sample are described in \cite[Matsuura et al. (2009)]{Matsuura09}, as well as other types of CMDs.
%\cite[Breysacher et al. (1999), Van Genderen (2001), Zickgraf (2006)]{Breysacher99, vanGenderen01, Zickgraf06}.
In mid-infrared CMDs, populous bright sources are AGB stars and red supergiants (RSGs), as well as some S\,Dor variables.
AGB isochrones from \cite[Marigo et al. (2008)]{Marigo08} are indicated here. Overall they trace 
observed distribution of AGB stars on the CMD quite well.

It has been suggested that AGB stars are important sources for $K$-band integrated flux of galaxies at age of 8--9 Gyr old.
Similarly, at the LMC, mid-infrared brightness of the integrated flux is attributed to AGB stars.
AGB stars contribute at least 20\,\% of the total flux of all point sources at 8.0\,$\mu$m.
This value is a lower limit, because we included AGB stars brighter than  [8.0]$<$12\,mag only, here;
below this magnitude, it is difficult to distinguish AGB stars and red-giants (RGs) only with this CMD.

{\underline{\it AGB mass-loss rate and its metallicity dependence}} 
 It has been an intensive debate on whether the mass-loss rates of AGB stars depends on the metallicities of the interstellar media (ISM) of galaxies, or not.
The mass-loss process of AGB stars involves dust grains surrounding the stars, and radiation pressure from the central stars on dust grains triggers the outflow.
The dust consists of metals, thus, the mass-loss rate could potentially be affected by the metallicities of galaxies.

%Among carbon-rich AGB stars,  `dust' mass-loss rate do not correlate with the metallcities of the ISM.
There is no observational evidence that  the `dust' mass-loss rate of carbon-rich AGB stars is related to the metallicities of the ISM, 
down to 5\,\% of the solar metallicity
\cite[(Matsuura et al. 2007; Sloan et al. 2009)]{Matsuura07, Sloan09}.
This is because AGB stars turn from oxygen-rich to carbon-rich due to carbon atoms synthesised in stars, and because
the mass of carbonaceous dust is strongly associated with self-produced carbon in AGB stars 
\cite[(Matsuura et al. 2002, 2005)]{Matsuura02, Matsuura05}.
There is an uncertainty caused by gas-to-dust mass ratio in the outflow. Nevertheless,
theoretical models predict that even the `gas' mass-loss rates of carbon-rich AGB stars are
 insensitive to the metallicities \cite[(Mattsson et al. 2008, Wachter et al. 2008)]{Mattsson08, Wachter08}.
Whereas, oxygen-rich AGB stars are expected to have a metallicity dependence of mass-loss rates \cite[(Habing 1996; Bowen \& Willson 1991)]{Habing96, Bowen96}.

 To incorporate the mass-loss rate into the stellar evolution models, we need to find the mass-loss rate as a function of stellar parameters,
 such as luminosities.
\cite[Groenewegen et al. (2007)]{Groenewegen07} showed the correlation between mass-loss rate and the period of stellar variability.
Periods can be converted to the `averaged' luminosities of the AGB stars \cite[(Feast et al. 1989)]{Feast89}; luminosities based on a single measurement
introduce more uncertainties.

Finally,  infrared colour is a direct probe of mass-loss rates for AGB stars.
Mass-loss rate of carbon-rich AGB stars follow a simple formula \cite[(Matsuura et al. 2009)]{Matsuura09},
and this relation is plotted at the top scale of Fig.\,\ref{fig1}.

{\underline{\it Age}} 
 To integrate the mass-loss rate into the stellar evolution model, the ages of the stars are a fundamental parameter.
In particular, high-mass (3$<M_{\rm ms}<$8$\,\msun$, where $M_{\rm ms}$ is the main sequence mass) 
AGB stars or high mass-loss rate ($>$10$^{-5}$\,\mlu) stars are relatively rare 
\cite[(Gruendl et al. 2008)]{Gruendl08} even in the field of the LMC. 
Nevertheless these stars are important for the chemical evolution of galaxies \cite[Matsuura et al. (2009)]{Matsuura09}.

 We have taken two sample of stars at the LMC bar center and north disc regions.
The analysis of the LMC star formation history suggests that these two places in the LMC have
experienced different periods in enhanced star formation rate \cite[(Cioni et al. 2006; Harris \& Zaritsky 2009)]{Cioni06, Harris09}.
The LMC bar has an enhanced population of 3--5 Gyrs old. The disc has mainly an older population ($>$11 Gyrs old),
although there are some young population (0.1 Gyr old) found in the disc.
Fig.\,\ref{fig2} shows the difference of CMDs at different locations (bar and disc) within the LMC.
The CMD of the bar contains very high mass-loss rate ($>3.5\times10^{-5}$\,\mlu) AGB stars, which is indicated as region (iv).
In contrast, there is no such a star in the LMC disc.
Region (iii), which includes  AGB stars with mass-loss rate of
$2\times10^{-7}$\,\mlu--$3.5\times10^{-5}$\,\mlu,  is more populous in the bar than the disc.
In region (ii), the disc population has much narrower distribution in the colour than the bar population.
Finally, region (i), which is dominated by emission objects, such as WR stars, are more populated in the disc.
These difference show that relatively high mass-loss AGB stars belong to 3--5 Gyrs (but $<$11\,Gyrs) old population.
%and not to the old population.

{\underline{\it Contribution to chemical enrichment in the ISM}} 
Table\,\ref{table1} shows the summary of gas and dust budget of the LMC.
This accounts for gas and dust injected from stars into the ISM.
%Details of the analysis is given in \cite[Matsuura et al. (2009)]{Matsuura09}.
This table shows that both AGB stars and supernovae (SNe) are important gas sources in the LMC.
Dust production in SNe remain largely uncertain, and AGB stars are one of the important dust sources.
More precise analysis of oxygen-rich AGB stars and RSGs will be forth coming.

{\underline{\it Summary}} 
Intensive data of the Local Group Galaxies have been taken by the SST, and their analysis is still on going. Before the JWST and the ELT era,
prescription of AGB mass loss will be largely improved.
Such analysis is important for galaxy from low-z to high-z galaxies \cite[(Sloan et al. 2009; Valiante et al. 2009)]{Sloan09}.

{\underline{\it Acknowledgement}} MM thanks to the IAU grant and Japan Astronomical Society for the financial support.

\end{document}